# Magnetic Susceptibility of Quasi-One-Dimensional Ising Superantiferromagnets FeTAC and $MCl_2 \cdot 2NC_5H_5$ (M = Co, Fe): Approximation with $L \times \infty$ and $L \times L \times \infty$ Chain Clusters

M. A. Yurishchev

*Institute of Problems of Chemical Physics, Russian Academy of Sciences,
Chernogolovka, Moscow oblast, 142432 Russia*
*e-mail: yur@itp.ac.ru*

**Abstract**—The temperature dependence of the zero-field susceptibilities of 2D and 3D Ising lattices with anisotropic coupling is analyzed. Infinite 2D and 3D lattices are approximated, respectively, by ensembles of independent $L \times \infty$ and $L \times L \times \infty$ chain clusters that are infinitely long in the strong-coupling ($J$) direction. This approach is used as a basis for a quantitative description of available experimental data on the magnetic susceptibilities of the 2D anisotropic Ising ferromagnet $[(CH_3)_3NH]FeCl_3 \cdot 2H_2O$ (FeTAC) and the quasi-one-dimensional 3D systems $CoCl_2 \cdot 2NC_5H_5$ and $FeCl_2 \cdot 2NC_5H_5$ in the entire experimental temperature range. A method is proposed for determining the relative interchain coupling strength $J'/J$ from the maximum susceptibility value, which improves the accuracy of estimates for $J'/J$ by more than an order of magnitude.

## 1. INTRODUCTION

Various currently known materials are Ising magnets [1, 2]. Even though the Ising coupling has an extremely simple form, the macroscopic properties of these materials, such as magnetic susceptibility, are very difficult to calculate. It should be noted that no rigorous closed-form expression for the longitudinal component of susceptibility has been obtained to this day even in exactly solvable Ising models. Moreover, there are reasons to believe that no such expression can be found in the class of differentially finite (holonomic) functions [3–5] (see also [6]).

The subject of this study is the magnetic susceptibility of quasi-one-dimensional Ising magnets. Systems of two types are considered: anisotropic square lattices with coupling constants $J$ and $J'$ such that $|J'/J| \leq 1$, and simple cubic lattices with dominant interaction along one axis represented by $J$ and equal constants $J'$ of interaction along the remaining two orthogonal axes.

Crystals of $[(CH_3)_3NH]FeCl_3 \cdot 2H_2O$ (FeTAC) have a 2D magnetic lattice consisting of bonded spin chains lying in a plane [7–10]. In crystals of $CoCl_2 \cdot 2NC_5H_5$ [11–14] and $FeCl_2 \cdot 2NC_5H_5$ [15–17], chains of magnetic ions are bonded into 3D systems. All of these materials are typical quasi-1D Ising superantiferromagnets that can be modeled by effective spin-1/2 Hamiltonians (with $J > 0$ and $J' < 0$). As temperature decreases, ferromagnetically ordered spin chains become antiferromagnetically ordered. Their magnetic susceptibilities have distinct maxima at temperatures $T_{max}$ above the respective critical points $T_c$. The phase transition manifests itself in the susceptibility curve as an inflection point where the tangent line to the curve is infinitely steep (in the ideal case).

The susceptibility of a 2D Ising lattice was calculated in [18] for the entire temperature range (in theory, from zero to infinity). The approximation used in that study (decoupling of many-spin correlation functions) is accurate within 0.35% in the isotropic model. However, the analysis presented below shows that the error in the coordinates of the susceptibility maximum amounts to tens of percent even for $J'/J = -0.1$ ($J > 0$). Therefore, this approximation cannot be applied to quasi-1D systems in practical calculations.

The results obtained for 3D systems are even less accurate. The most reliable calculations of susceptibility for such systems make use of power series expansions. For the zero-field longitudinal susceptibility of the isotropic simple cubic Ising lattice, high-temperature expansions to the 25th- and even 32th-order terms were obtained in [19, 20] and [21], respectively. However, analogous expansions for anisotropic lattices are known only to the 10th- or 11th-order terms (see [22] and [23], respectively). Moreover, partial sums of the series rapidly diverge with increasing lattice anisotropy. In what follows, it is demonstrated that the available high-temperature series expansions of superantiferromagnetic susceptibility [24] result in unacceptably large errors for $|J'|/J = 10^{-2}$ (even after their conver-gence is improved by Padé–Borel resummation). Note that interpretation of the experimental data discussed here requires modeling with an even smaller value of this parameter.

In this paper, susceptibilities are calculated by using cluster simulation. It is well known [25–27] that various characteristics calculated by this method (including susceptibility) converge to their values for an infinite system at an exponential rate with increasing cluster size everywhere in the parameter space except for a narrow critical region. However, this region cannot be resolved by modern experimental methods for the quasi-1D materials discussed here.

In view of the specific anisotropy to be modeled, chain clusters of infinite length in the direction of the dominant interaction $J$ are used as subsystems ($L \times \infty$ strips and $L \times L \times \infty$ parallelepipeds for 2D and 3D systems, respectively). Undesirable surface effects are eliminated by setting periodic boundary conditions in the transverse directions for subsystems of both types. Furthermore, frustration is obviated by using chains of length $L = 2, 4, \ldots$ (measured in units of the lattice constant), with the only exception of a single chain ($L = 1$). Thus, the magnetic lattice of an $L^{d-1} \times \infty$ superantiferromagnetic cluster (in space of dimension $d = 2$ or 3) consists of two identical interpenetrating sublattices with opposite magnetic moments.

In Section 2, formulas for susceptibilities are presented, including both general expressions well suited for computations and exact asymptotic ones. The cumbersome analytical formulas derived for few-chain subsystems are relegated to the Appendix. In Section 3, the strip width ensuring the accuracy required to calculate the susceptibility of FeTAC is determined. In Section 4, the corresponding calculated results are presented. Sections 5 and 6 contain results for 3D systems analogous to those presented in the preceding two sections. Section 7 summarizes the principal results of this study.

## 2. CALCULATION OF SUSCEPTIBILITIES

The anisotropic Ising Hamiltonian is written as

$$H = -\frac{1}{2}J \sum_{\langle i,j \rangle} \sigma_i^z \sigma_j^z - \frac{1}{2}J' \sum_{[i,j]} \sigma_i^z \sigma_j^z, \quad (1)$$

where the Pauli matrices $\sigma_i^z$ are localized at the sites of a square or simple cubic lattice. The sums with $\langle i,j \rangle$ and $[i,j]$ are taken over the nearest-neighbor pairs along the directions corresponding to $J$ and $J'$, respectively.

According to Kubo's linear response theory [28, 29], the static zero-field susceptibility tensor is

$$\chi_{\mu\nu} = -\beta \langle M_\nu \rangle \langle M_\mu \rangle + \int_0^\beta d\beta' \langle M_\nu(\beta') M_\mu \rangle. \quad (2)$$

Here, $\mu$ and $\nu$ stand for $x$, $y$, or $z$; $\beta = 1/k_B T$ is the inverse temperature measured in energy units ($k_B$ is Boltzmann's constant); angle brackets denote ensemble-averaged quantities; $M_\mu$ is the projection of the magnetic moment of the system on the $\mu$ axis; and $M_\nu(\beta) = e^{\beta H} M_\nu e^{-\beta H}$ is a component of magnetization in the Matsubara representation.

Note that superantiferromagnets, being characterized by zero total spontaneous sublattice magnetizations, have zero magnetic moments in the absence of applied field: $\langle M_\nu \rangle = 0$. Therefore, the first term on the right-hand side of (2) vanishes under the conditions considered in this study.

The component of the magnetic moment parallel to the $z$ axis is

$$M_z = \frac{1}{2} g_\parallel \mu_B \sum_{i=1}^{N} \sigma_i^z, \quad (3)$$

where $g_\parallel$ is the longitudinal $g$ factor, $\mu_B$ is the Bohr magneton, and $N$ is the total number of particles in the system. Since $M_z$ commutes with Hamiltonian (1) and $\langle M_z \rangle = 0$, the expression for the molar zero-field longitudinal (parallel) susceptibility obtained by substituting (3) into (2) is

$$\chi_\parallel(T) \equiv \lim_{N \to \infty} \frac{N_A}{N} \chi_{zz} = \frac{N_A g_\parallel^2 \mu_B^2}{4 k_B T} \lim_{N \to \infty} \sum_{j=1}^{N} \langle \sigma_{i_0}^z \sigma_j^z \rangle_N, \quad (4)$$

where $N_A$ is Avogadro's number and $i_0$ is any particular site in a uniform lattice ($\chi_\parallel$ is independent of its location). To evaluate the longitudinal susceptibility, one must calculate and add up all spin–spin correlation functions and take the infinite-lattice limit.

The longitudinal susceptibilities of single-, double-, and four-chain Ising models are known in analytical form (see Appendix). An analysis of these formulas shows that the predicted variation of the susceptibilities of superantiferromagnetic clusters with temperature is in qualitative agreement with experimental data. The susceptibility curve has a peak (see Fig. 1), and its magnitude indefinitely increases with lattice anisotropy. At temperatures below the maximum point, the susceptibility curve has an inflection point that approximately corresponds to the critical point of the entire system. The slope of the tangent line at the inflection point increases with the number of chains in a subsystem, approaching infinity.

For subsystems consisting of a larger number of chains, the susceptibility can be found only by numerical methods. One formula well suited for computing the longitudinal susceptibility of an $L^{d-1} \times \infty$ Ising cluster by the transfer matrix method is [30, 31]

$$\chi_\|^{(L^{d-1} \times \infty)}(T) = \frac{N_A g_\|^2 \mu_B^2}{4L^{d-1} k_B T} \\ \times \sum_s{}' \frac{\lambda_1 + \lambda_s}{\lambda_1 - \lambda_s} |f_s^+ (\hat{\sigma}_1 + \ldots + \hat{\sigma}_n) f_1|^2. \quad (5)$$

Here, the primed sum skips the term with $s = 1$; $n = L^{d-1}$ is the number of chains in a cluster; $\lambda_s$ and $f_s$ denote eigenvalues and the corresponding eigenvectors of the transfer matrix, respectively (the largest eigenvalue $\lambda_1$ is nondegenerate by the Perron–Frobenius theorem [32]); and the matrix $\hat{\sigma}_k$ is defined as

$$\hat{\sigma}_k = 1 \times \ldots \times 1 \times \sigma_z \times 1 \times \ldots \times 1. \quad (6)$$

In this direct product of $n$ matrices, the $k$th multiplicand is the Pauli matrix $\sigma_z$, and the remaining ones are two-dimensional identity matrices. The $2 \times 2^n$-by-$2 \times 2$ transfer matrix $V$ has the elements

$$\langle \sigma_1, \ldots, \sigma_n | V | \sigma'_1, \ldots, \sigma'_n \rangle \\ = \exp\left[ K \sum_{i=1}^n \sigma_i \sigma'_i + \frac{1}{2} K' \sum_{[i,j]} (\sigma_i \sigma_j + \sigma'_i \sigma'_j) \right], \quad (7)$$

where $\sigma_i = \pm 1$ are collinear spins in the cross section of an $L^{d-1} \times \infty$ lattice, $K = J/2k_B T$, and $K' = J'/2k_B T$. The transfer matrix $V$ is a positive real symmetric one.

Expression (5) follows from (4); i.e., it can be derived from Kubo's linear response theory. A physically equivalent expression that has a somewhat different form was obtained by developing a perturbation series in external field for the transfer matrix [33, 34].

The key problem in evaluating the susceptibility is thus reduced to the eigenvalue–eigenvector problem for the transfer matrix $V$. Here, this problem is solved either by direct numerical diagonalization of the matrix $V$ or by diagonalizing the subblocks constituting the transfer matrix block diagonalized by using cluster symmetries.

Starting again from Kubo's formula (2), one can readily show that the expression for the transverse susceptibility $\chi_\perp$ is a linear combination of a finite number of local $\sigma^z$ correlation functions. These correlation functions and $\chi_\perp$ have been calculated only for isotropic 2D Ising lattices [35–37].

The available analytical expressions for the transverse susceptibilities of single- and double-chain Ising models are written out in the Appendix. Figure 2 shows these susceptibilities as functions of temperature and demonstrates that the transverse susceptibility of a quasi-2D system only slightly deviates from that of a single chain in the entire temperature range, in contrast to the longitudinal susceptibility. Therefore, practical calculations of the transverse susceptibility of a typical quasi-2D magnet can be performed by using the formula for a single chain.

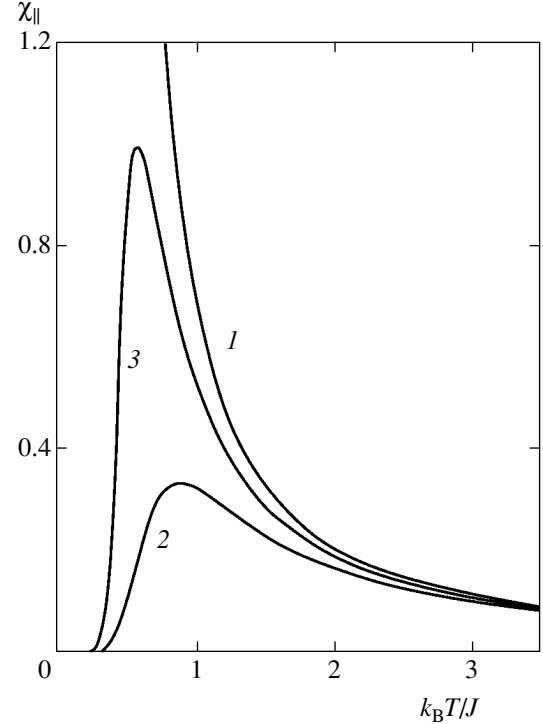

**Fig. 1.** Longitudinal susceptibility for Ising chain models (measured in units of $N_A g_\|^2 \mu_B^2 /J$) vs. normalized temperature: (*1*) linear ferromagnetic chain; (*2*) double chain with $z'J'/J = -0.5$ ($J > 0$); (*3*) four-chain cylinder with $J > 0$ and $z'J'/2J = -0.1$.

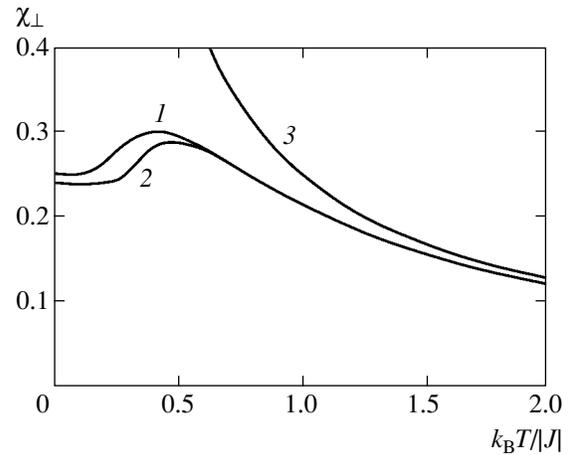

**Fig. 2.** Zero-field transverse susceptibility (measured in units of $N_A g_\perp^2 \mu_B^2 /J$) vs. normalized temperature: (*1*) 1D chain ($J' = 0$); (*2*) double chain with $z'|J'/J| = 0.1$; (*3*) free spins ($J = J' = 0$).

An Ising magnet is an easy-axis one. The susceptibility of such a material in polycrystalline (powder) form is

$$\chi_p = \frac{1}{3}\chi_\parallel + \frac{2}{3}\chi_\perp. \quad (8)$$

At high temperatures, the longitudinal susceptibility of an Ising magnet obeys the Curie–Weiss law:

$$\chi_\parallel(T) \approx \frac{C_\parallel}{T - \vartheta}, \quad T \longrightarrow \infty. \quad (9)$$

This expression follows from the lowest order terms in the high-temperature expansion of the susceptibility of an infinite Ising lattice with anisotropic coupling [22, 23]. The Curie constant in (9) is expressed as

$$C_\parallel = \frac{N_A g_\parallel^2 \mu_B^2}{4 k_B}, \quad (10)$$

and the Curie–Weiss temperature is

$$\vartheta = \frac{1}{k_B}\left(J + \frac{1}{2}z'J'\right). \quad (11)$$

Here, $z'$ is the number of nearest-neighbor chains in the system: $z' = 2$ and 4 for 2D and 3D lattices, respectively.

It follows from exact low-temperature expansions obtained in [36, 38, 39] that the longitudinal susceptibility of the entire system in the two- or higher dimensional space vanishes exponentially as $T \longrightarrow 0$. At absolute zero temperature,

$$\chi_\parallel(0) = 0. \quad (12)$$

Since the clusters with even $L$ considered here have zero magnetic moments and an infinitesimal external field cannot induce any magnetic moment that requires a finite amount of work to be done, boundary condition (12) is automatically satisfied for $L^{d-1} \times \infty$ subsystems with $d > 1$.

At high temperatures, the transverse susceptibility obeys the Curie law [36, 40]:

$$\chi_\perp(T) \approx \frac{C_\perp}{T}, \quad T \longrightarrow \infty. \quad (13)$$

(It is equal to the magnetic susceptibility of free spins.) The Curie constant in (13) is expressed as

$$C_\perp = \frac{N_A g_\perp^2 \mu_B^2}{4 k_B}, \quad (14)$$

where $g_\perp$ is the transverse $g$ factor. As $T \longrightarrow 0$, the transverse susceptibility tends to a finite limit [35, 36] (see also [40]). For the anisotropic lattices of interest for the present study,

$$\chi_\perp(0) = \frac{N_A g_\perp^2 \mu_B^2}{2(2|J| + z'|J'|)}. \quad (15)$$

An expression for the relative interchain coupling strength is found by combining (11), (14), and (15):

$$\frac{|J'|}{J} = \frac{z'}{2}\frac{C_\perp - \vartheta\chi_\perp(0)}{C_\perp + \vartheta\chi_\perp(0)}. \quad (16)$$

For $|J'|/J > 0$, it entails the constraint

$$\chi_\perp(0) < \frac{C_\perp}{\vartheta}. \quad (17)$$

Thus, the susceptibility measured at high temperatures imposes a constraint on its value at extremely low temperatures. A comparison shows that the experimental data obtained for FeTAC in [7] satisfy inequality (17).

According to (8)–(11), (13), and (14), the powder susceptibility at high temperatures also decreases as

$$\chi_p(T) \approx \frac{C}{T - \Theta} \quad (T \longrightarrow \infty), \quad (18)$$

where

$$C = \frac{N_A(g_\parallel^2 + 2g_\perp^2)\mu_B^2}{12 k_B}, \quad (19)$$

$$\Theta = \frac{g_\parallel^2}{(g_\parallel^2 + 2g_\perp^2)k_B}\left(J + \frac{1}{z'}z'J'\right). \quad (20)$$

At zero temperature, the powder susceptibility reduces to its transverse component:

$$\chi_p(0) = \frac{N_A g_\perp^2 \mu_B^2}{3(2|J| + z'|J'|)}. \quad (21)$$

Expressions (19)–(21) combined with the inequality $(g_\parallel^2 - 2g_\perp^2)^2 \geq 0$ yield a constraint on the relative coupling strength in an anisotropic system:

$$\left|\frac{J'}{J}\right| \leq \frac{2}{z'}\frac{C - 4\Theta\chi_p(0)}{C + 4\Theta\chi_p(0)}. \quad (22)$$

This inequality, in turn, entails an upper bound for the low-temperature plateau in powder susceptibility with $\Theta > 0$:

$$\chi_p(0) < \frac{C}{4\Theta}. \quad (23)$$

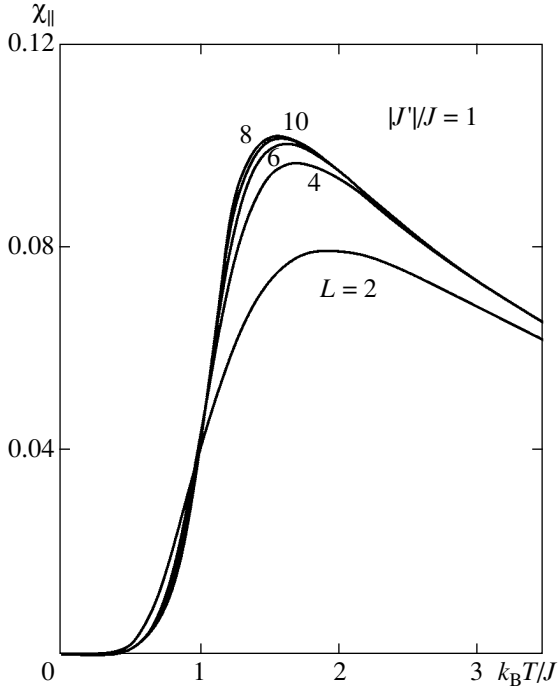

**Fig. 3.** Longitudinal susceptibility for $L \times \infty$ superantiferromagnetic Ising strips with $L = 2$–$10$ and $|J'|/J = 1$ (measured in units of $N_A g_\parallel^2 \mu_B^2 /J$).

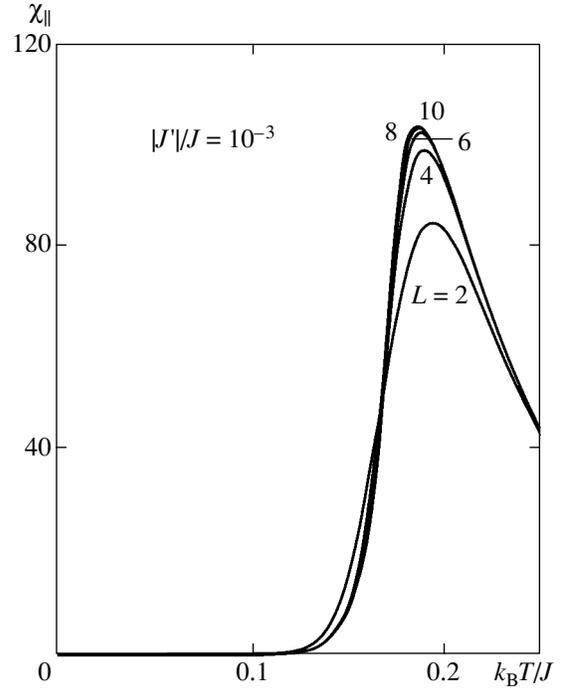

**Fig. 4.** Longitudinal susceptibility for $L \times \infty$ superantiferromagnetic Ising strips with $L = 2$–$10$ and $|J'|/J = 10^{-3}$ (measured in units of $N_A g_\parallel^2 \mu_B^2 /J$).

It is shown in Section 6 that available experimental data for $CoCl_2 \cdot 2NC_5H_5$ [11] support this inequality.

## 3. CONVERGENCE ANALYSIS OF CLUSTER EXPANSIONS FOR 2D SYSTEMS

Before applying any particular approximation to quantify experimental data, it should be verified that the systematic error of the approximation is smaller (at least, not greater) than the measurement errors.

First, let us analyze the convergence of cluster expansions for 2D systems. Figures 3 and 4 show the longitudinal susceptibilities of $L \times \infty$ Ising strips with $L$ varying from 2 to 10 computed for $|J'|/J = 1$ and $10^{-3}$, respectively. The interval between these extreme values contains the relative interchain coupling strengths characteristic of most Ising magnets actually used in experiments. When $|J'|/J$ is smaller, dipole–dipole interaction plays a significant role.

The susceptibilities were computed by using formula (5), where the eigenvalues and eigenvectors were found by direct numerical diagonalization of the starting transfer matrices having dimensions no higher than $2^{10} = 1024$ with the use of the C subroutines *tred2* and *tqli* [41]. The output data also included the coordinates of the susceptibility maximum.

Figures 3 and 4 illustrate the convergence of susceptibility with increasing subsystem size. In both extreme cases, $|J'|/J = 1$ and $10^{-3}$, the curves obtained for the strips with $L = 8$ and 10 nearly coincide; i.e., these results correspond to the infinite 2D lattice up to the resolution of the graphs.

The approximation accuracy can be reliably estimated by comparing the maximum values of longitudinal susceptibility. The maximum is in the subcritical region, where the convergence follows a stretched exponential, if not a power, law.

Table 1 lists the coordinates of the longitudinal susceptibility maxima for Ising cylinders. The extrapolation to the thermodynamic limit ($L = \infty$) was performed by applying the Shanks transform [26, p. 225], which maps a sequence $\{a_l\}$ to $\{a_l'\}$ according to the formula

$$a_l' = \frac{a_{l-1} a_{l+1} - a_l^2}{a_{l-1} + a_{l+1} - 2a_l}. \quad (24)$$

The results presented in Table 1 demonstrate the following trends. The relative estimation error decreases with weakening interchain coupling from 1.3% for $|J'|/J = 1$ to 0.27% for $|J'|/J = 10^{-3}$. Furthermore, the maximum longitudinal susceptibility increases with decreasing $|J'|/J$ as

$$\frac{J \chi_\parallel^{(\max)}}{N_A g_\parallel^2 \mu_B^2} \approx \frac{0.1}{|J'|/J}. \quad (25)$$

**Table 1.** Coordinates of longitudinal-susceptibility maxima for cyclic $L \times \infty$ superantiferromagnetic Ising strips with different $L$ and $\Delta = |J'|/J$ ($J > 0$, $J' < 0$): upper and lower values are $k_B T_{max}/J$ and $J\chi_{\parallel, max}^{(L \times \infty)}/N_A g_\parallel^2 \mu_B^2$, respectively. Extrapolation to 2D strips with $L = \infty$ is performed by applying Shanks transform (24) to strips with $L = 6$, 8, and 10.

| $L$ | $\Delta = 1$ | $\Delta = 10^{-1}$ | $\Delta = 10^{-2}$ | $\Delta = 10^{-3}$ |
|---|---|---|---|---|
| 2 | 1.957443 | 0.593380 | 0.301827 | 0.193399 |
|   | 0.080142 | 0.853605 | 8.577071 | 85.786055 |
| 4 | 1.723455 | 0.563281 | 0.292302 | 0.189102 |
|   | 0.097244 | 1.000137 | 10.019720 | 100.203811 |
| 6 | 1.640401 | 0.550014 | 0.287992 | 0.187135 |
|   | 0.101008 | 1.034448 | 10.360714 | 103.612905 |
| 8 | 1.603729 | 0.543665 | 0.285905 | 0.186177 |
|   | 0.102075 | 1.044801 | 10.463915 | 104.644760 |
| 10 | 1.587364 | 0.540669 | 0.284914 | 0.185721 |
|    | 0.102404 | 1.048171 | 10.497584 | 104.981419 |
| $\infty$ | 1.57(2) | 0.538(3) | 0.2840(9) | 0.1853(5) |
|          | 0.1026(2) | 1.050(2) | 10.51(2) | 105.1(2) |

However, the error of estimation of the maximum value, unlike $k_B T_{max}/J$, is almost independent of lattice anisotropy. For $L = 10$, it is approximately 0.2%. This accuracy is sufficient for quantitative description of available experimental data.

Let us now discuss the accuracy of the approximation used in [18]. According to Table 1, we find that the maximum longitudinal susceptibility of the 2D lattice with $|J'|/J = 10^{-1}$ corresponds to $k_B T_{max}/J = 0.538(3)$, and its value is $J\chi_\parallel^{(max)}/N_A g_\parallel^2 \mu_B^2 = 1.050(2)$. On the other hand, the results presented in [18, Fig. 5] for the same superantiferromagnet demonstrate that the maximum has the coordinates $k_B T_{max}/J \approx 0.656$ and $J\chi_\parallel^{(max)} \approx 0.809$ (with redefined coupling constants). Thus, the errors of estimation of the maximum value and the corresponding temperature in the approximation used in [18] are 22 and 23%, respectively; i.e., the theory developed in [18] is too inaccurate to be applicable to experimental data even for weakly anisotropic lattices.

The inverse variation of maximum longitudinal susceptibility of superantiferromagnetic lattices with relative interchain coupling strength described by empirical formula (25) can be explained as follows. The critical temperature for anisotropic 2D Ising lattice satisfies the equation [42]

$$\sinh\frac{|J|}{k_B T_c} \sinh\frac{|J'|}{k_B T_c} = 1. \quad (26)$$

As $|J'|/J \longrightarrow 0$, this equation yields

$$\frac{|J'|}{J} \approx 2\frac{k_B T_c}{J}\exp\left(-\frac{J}{k_B T_c}\right). \quad (27)$$

On the other hand, the longitudinal susceptibility of a typical quasi-1D Ising superantiferromagnet at temperatures slightly above the maximum point is well approximated by the formula for the longitudinal susceptibility of a single Ising chain. Moreover, the maximum point approaches the critical temperature with increasing lattice anisotropy. Therefore, assuming that $T_{max} \approx T_c$ and $\chi_\parallel^{(max)} \propto \chi_\parallel^{(1D)}(T_{max})$, we combine (27) with (A.1) to obtain

$$\frac{J\chi_\parallel^{(max)}}{N_A g_\parallel^2 \mu_B^2} \propto \left(\frac{|J'|}{J}\right)^{-1}. \quad (28)$$

The data listed in Table 1 demonstrate that relation (28) holds in a surprisingly wide interval extending almost to $|J'| = J$.

## 4. ZERO-FIELD MAGNETIC SUSCEPTIBILITY OF SINGLE-CRYSTAL FeTAC

Single crystals of ferrous trimethylammonium chloride (FeTAC) are characterized by the most pronounced quasi-one-dimensional magnetic ordering among the class of compounds described by the formula $[(CH_3)_3NH]MX_3 \cdot 2H_2O$, where M denotes a metal (such as Co, Fe, or Ni) and X is chlorine or bromine (see [8, 43] and references therein). Physical properties of FeTAC are the subject of extensive experimental studies.

The static zero-field magnetic susceptibility of FeTAC single crystals was measured in [7] at temperatures ranging from 1.4 to 300 K. The susceptibility along the easy axis (crystallographic $b$ axis) is interpreted as the longitudinal susceptibility: $\chi_b \equiv \chi_\parallel$. According to [7], its maximum value $\chi_b^{(max)} = 100$ cm³/mol is reached at $T_{max} = 3.18(2)$ K, and the critical temperature determined from the steepest slope of longitudinal susceptibility below $T_{max}$ is $T_c = 3.12(2)$ K (i.e., $T_{max}/T_c \approx 1.02$). At temperatures well above $T_{max}$, the susceptibility obeys the Curie–Weiss law.

The quantitative interpretation of measured susceptibilities presented in [7] is based on the single-chain approximation. Therefore, it is applicable only at temperatures above $T_{max}$. By fitting the longitudinal susceptibility of the 1D Ising chain to experimental data points in the interval between 6 and 18 K, it was found that $C_b = C_\parallel = 5.52(4)$ cm³ K/mol (Curie constant) and $\vartheta \approx J/k_B = 16.6(1)$ K [7]. Combining these results with (10), we obtain the longitudinal $g$ factor: $g_\parallel = 7.67(3)$.

Moreover, the relative interchain coupling strength was estimated in [7] by using a well-known expression for the susceptibility of a quasi-1D system with coupling between chains described in the molecular field approximation [44],

$$\chi_\parallel^{(MFA)} = \chi_\parallel^{(1D)} \Big/ \left(1 - \frac{z'J'}{2C_\parallel}\chi_\parallel^{(1D)}\right). \quad (29)$$

By fitting this theoretical formula to experimental data on susceptibility in the interval between 3.2 K (again above $T_{max}$) and 18 K, the approximate value $J'/J \approx -2 \times 10^{-3}$ was obtained [7].

According to the results of measurements of the specific heat of FeTAC reported in [8], $T_c = 3.125(5)$ K and $J/k_B = 17.7(3)$ K. Onsager's solution (26) was used in [8] to obtain $|J'/J| = 1.3 \times 10^{-3}$ for FeTAC.

Before discussing the results on FeTAC obtained in the present study, let us note that the use of dependence of $k_B T_c/J$ on $J'/J$ is not the best method for finding the interchain coupling strength in quasi-1D systems. Indeed, (27) entails the following relation between the relative errors in normalized interchain coupling and reduced temperature:

$$\delta\!\left(\frac{|J'|}{J}\right) = \left[1 + \left(\frac{k_B T_c}{J}\right)^{-1}\right]\delta\!\left(\frac{k_B T_c}{J}\right). \quad (30)$$

Therefore, the error in $|J'|/J$ estimated from $T_c$ increases with coupling anisotropy (since $k_B T_c/J$ decreases).

The same conclusion can be reached in a different manner. Transcendental equation (26) is easily solved on a computer to obtain curve 1 in Fig. 5. However, the inaccuracy of input data should be taken into account. Following [8], let us use $T_c = 3.125 \pm 0.005$ K and $J/k_B = 17.7 \pm 0.3$ K. Then, $k_B T_c/J = 0.177 \pm 0.003$, the corresponding relative error is 1.7%, and the relative error in the result $|J'|/J = (1.1-1.4) \times 10^{-3}$ obtained by solving (26) is 12%. This sharp increase in error is explained by a rapid increase in the steepness of curve 1 with decreasing $J'/J$ (see Fig. 5).

Alternatively, the ratio $|J'|/J$ can be determined for a superferromagnetic system by using maximum susceptibility values (curve 2 in Fig. 5). According to (25), $J\chi_\parallel^{(max)}/N_A g_\parallel^2 \mu_B^2$ varies in inverse proportion to $|J'|/J$. This relation obviously implies that the respective relative errors in these parameters are equal; therefore,

$$\delta\!\left(\frac{|J'|}{J}\right) = \delta(\chi_\parallel^{(max)}) + \delta\!\left(\frac{J}{k_B}\right) + \delta(C_\parallel). \quad (31)$$

One important advantage of this method for estimating $|J'|/J$ over the one discussed above is that the error is independent of lattice anisotropy.

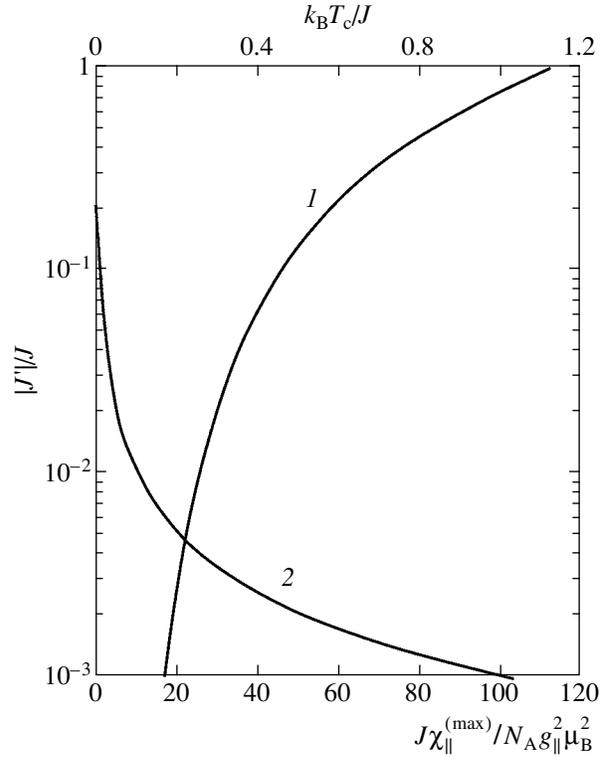

**Fig. 5.** Reduced critical temperature (curve 1) and maximum value of normalized longitudinal susceptibility (curve 2) vs. $|J'|/J$ for 2D superantiferromagnetic Ising lattices.

The principal result of this section is a quantitative description of the magnetic susceptibility of FeTAC in the entire temperature range down to absolute zero. Computations were performed for a 10-chain strip to ensure that the approximation error is much smaller than measurement errors. ($L \times \infty$ strips with $L = 12$, 14, and larger could easily be simulated on a modern computer if necessary.)

Expression (5) for longitudinal susceptibility contains three parameters: $J/k_B$, $g_\parallel$, and $J'/J$. As in [7], the function $\chi_\parallel^{(1D)}(T)$ was fitted to experimental data points at temperatures of 6 K and higher to obtain $J/k_B = 16.6$ K and $g_\parallel = 7.67$. By matching the computed maximum susceptibility with that measured in [7] ($T_{max} = 3.18$ K, $\chi_\parallel^{(max)} = 100$ cm$^3$/mol), it was found that $J'/J = -0.00138$ with an error in the last digit.

Figure 6 demonstrates that the curve computed in the present study agrees with experimental data for FeTAC obtained in [7]. It should be reiterated here that the theoretical description of experimental data on susceptibility given in [7] is valid only at temperatures above the maximum point, when the single-chain approximation is applicable.

## 5. CLUSTER EXPANSIONS FOR 3D SYSTEMS

Modern computers can be used to simulate 3D Ising chain clusters only for $L \leq 4$. Note that exact expressions are available for the susceptibilities of the double-chain cluster and the $2 \times 2 \times \infty$ parallelepiped (see Appendix).

To find the eigenvalues of the $65\,536 \times 65\,536$ transfer matrix of the cyclic $4 \times 4 \times \infty$ Ising parallelepiped, both lattice and spin symmetries were used to represent it in a block diagonal form consisting of $433 \times 433$ and $372 \times 372$ subblocks, whose eigenvalues and eigenvectors are required to calculate the susceptibility given by (5) [34]. Exact diagonalization of these relatively small subblocks can readily be performed on a PC.

The next larger cluster that should have been simulated in the present study is the $6 \times 6 \times \infty$ parallelepiped. However, the corresponding transfer matrix is $2^{36} \times 2^{36}$, and the dimensions of the subblocks in its block diagonal form determined by using the symmetries of the system (as in the case of the $4 \times 4 \times \infty$ cluster) are $119\,583\,470$ and $119\,539\,680$ [34]. The complete solution of the spectral problem for these matrices is far beyond the capabilities of present-day supercomputers. To date, computations have been performed for the $6 \times 6 \times \infty$ Ising system only at the quantum limit and a few lowest eigenvalues of the corresponding sparse Hamiltonian matrix have been calculated [45].

For this reason, the present analysis is restricted to $2 \times \infty$, $2 \times 2 \times \infty$, and $4 \times 4 \times \infty$ clusters. Figure 7 illustrates the convergence of the zero-field susceptibilities computed for 3D clusters (curves 2–4) with respect to the number of chains in a cluster. The trends shown here are qualitatively similar to those manifested in the 2D simulations (Figs. 3 and 4): the maximum susceptibility increases with the cluster size, while the corresponding reduced temperature decreases. As $L \longrightarrow \infty$, the coordinates of the maximum must approach their respective limits. However, these limits cannot be calculated by Shanks extrapolation for lack of solution for the $6 \times 6 \times \infty$ cluster. (Unfortunately, the $2 \times \infty$ chain cannot be used in an extrapolation process, because it is not a truly 3D cluster.)

The accuracy of the $4 \times 4 \times \infty$ approximation used here for comparison with experiment can be estimated indirectly by invoking the results of a Padé–Borel resummation of high-temperature expansions for the susceptibility of an anisotropic 3D Ising lattice [24]. In that study, the coordinates of the superantiferromagnetic susceptibility maximum were presented as func-tions of $J'/J$ (see Table 1 in [24]). In the isotropic case ($|J'|/J = 1$), when high-temperature expansions must lead to reliable results, it was found that $k_B T_{max}/J = 2.400(25)$ and $J\chi_\parallel^{(max)}/N_A g_\parallel^2 \mu_B^2 = 0.058$ [24]. In this study, $J\chi_{\parallel, max}^{(4 \times 4 \times \infty)}/N_A g_\parallel^2 \mu_B^2 = 0.05547$ is obtained for the $4 \times 4 \times \infty$ lattice, which is lower by 4.35%. By analogy with the 2D simulations, it can be assumed that the relative error in the maximum value is independent of lattice anisotropy. (Actually, the error even slightly decreases with $|J'|/J$, which can easily be demonstrated by using the results for $L = 4$ presented in [24, Table 1].) Then, the error of the $4 \times 4 \times \infty$ approximation must be about 4%. Since theoretical results are compared here with experimental data on powder susceptibility, which are not very accurate, the $4 \times 4 \times \infty$ cluster approximation is well suited for such a comparison.

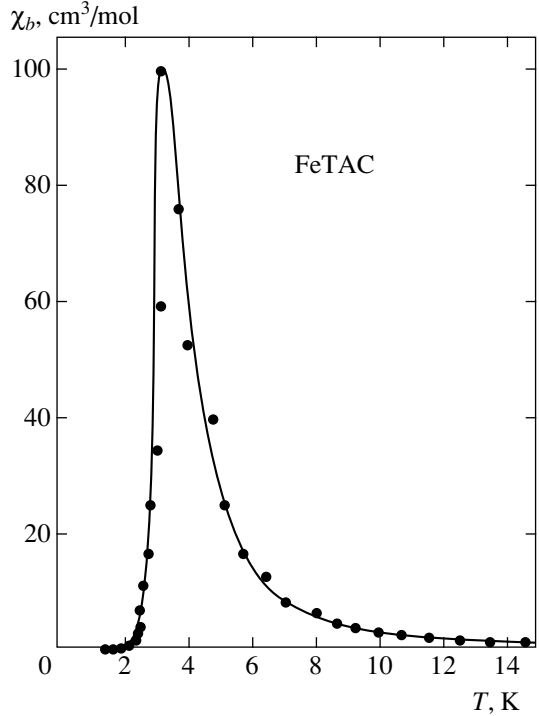

**Fig. 6.** Magnetic susceptibility of FeTAC along the crystallographic $b$ axis: measurement data from [7] (symbols) and $\chi_\parallel^{(10 \times \infty)}(T)$ calculated for $J/k_B = 16.6$ K, $g_\parallel = 7.67$, and $J'/J = -1.38 \times 10^{-3}$ (curve).

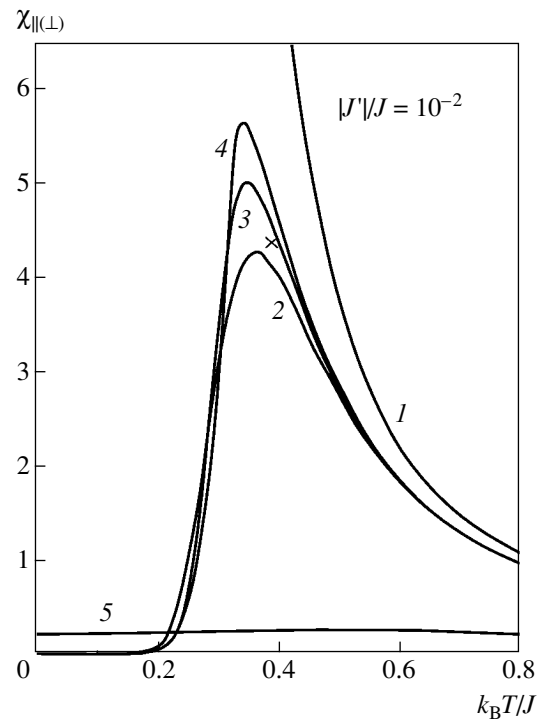

**Fig. 7.** Longitudinal susceptibility for Ising clusters with $|J'|/J = 10^{-2}$ (measured in units of $N_A g_\parallel^2 \mu_B^2 /J$): (1) linear ferromagnetic chain; (2) double chain with quadrupled interchain coupling; (3) cyclic $2 \times 2 \times \infty$ and (4) cyclic $4 \times 4 \times \infty$ parallelepipeds. The cross is the maximum calculated in [24]. The transverse susceptibility of a linear Ising chain (measured in units of $N_A g_\perp^2 \mu_B^2 /J$) is shown for comparison (curve 5).

As mentioned in the Introduction, high-temperature expansions lead to increasingly inaccurate results with decreasing $|J'|/J$, because they contain a small number of terms. The cross in Fig. 7 represents the susceptibility maximum calculated in [24] for $|J'|/J = 10^{-2}$, with $k_B T/J = 0.385(5)$ and $J\chi_\parallel/N_A g_\parallel^2 \mu_B^2 = 4.40(75)$. According to the figure, high-temperature expansions are even less reliable than the $2 \times 2 \times \infty$ approximation for this degree of anisotropy. These results cannot be compared to experimental data for such highly anisotropic superantiferromagnets as $CoCl_2 \cdot 2NC_5H_5$ (with $|J'|/J < 10^{-2}$). Now, it is clear that the formal quantitative comparison of this kind presented in [24] is groundless.

Table 2 summarizes the coordinates of the susceptibility maximum calculated for the $4 \times 4 \times \infty$ superantiferromagnetic cluster. (The maximum longitudinal susceptibility is divided by 3 with a view to comparing with powder susceptibility.) As in the case of 2D lattice, the maximum longitudinal susceptibility varies in inverse proportion with $|J'|/J$:

$$\frac{J\chi_{\parallel, max}^{(4 \times 4 \times \infty)}}{3 N_A g_\parallel^2 \mu_B^2} \approx \frac{0.019}{|J'|/J}. \quad (32)$$

This behavior is explained by analogy with the 2D case: when the coupling between chains is described in the molecular field approximation (while the intrachain coupling is modeled exactly), the critical temperature for the ferromagnetic simple cubic Ising lattice satisfies the transcendental equation [44]

$$K_B T_c = \frac{z'}{2} J' \exp\left(\frac{J}{k_B T_c}\right). \quad (33)$$

Since $T_{max} \approx T_c$ and the maximum susceptibility is on the order of $\chi_\parallel^{(1D)}(T_{max})$ for anisotropic superantiferromagnetic lattices with sufficiently high anisotropy, the law formulated in (28) follows again from Eq. (A.1).

**Table 2.** Normalized coordinates of longitudinal-susceptibility maximum as a function of $J'/J$ for $4 \times 4 \times \infty$ lattices with $J > 0$ and $J' < 0$

| $J'/J$ | $k_B T_{max}/J$ | $\frac{1}{3}\chi_{\parallel, max}^{(4 \times 4 \times \infty)}$ |
|---|---|---|
| −1.0000 | 2.5833625 | 0.01849229 |
| −0.1000 | 0.7155626 | 0.18871242 |
| −0.0100 | 0.3399309 | 1.89015328 |
| −0.0095 | 0.3346111 | 1.98964522 |
| −0.0090 | 0.3301889 | 2.10018741 |
| −0.0085 | 0.3254978 | 2.22374800 |
| −0.0080 | 0.3207257 | 2.36273906 |
| −0.0075 | 0.3158513 | 2.52025948 |
| −0.0070 | 0.3105331 | 2.70029750 |
| −0.0065 | 0.3052407 | 2.90802557 |
| −0.0060 | 0.2996506 | 3.15038075 |
| −0.0055 | 0.2936108 | 3.43680225 |
| −0.0050 | 0.2875449 | 3.78048335 |
| −0.0045 | 0.2808194 | 4.20055926 |
| −0.0040 | 0.2738453 | 4.72563875 |
| −0.0035 | 0.2661257 | 5.40079611 |
| −0.0030 | 0.2578018 | 6.30088685 |
| −0.0025 | 0.2483499 | 7.56111673 |
| −0.0020 | 0.2378198 | 9.45146198 |
| −0.0015 | 0.2252555 | 12.6019164 |
| −0.0010 | 0.2095811 | 18.9030324 |

Table 3 lists the coordinates of the inflection point below the maximum and the corresponding slopes of the susceptibility curve. It is clear that the abscissa of the inflection point is a lower bound for critical temperature (cf. [34]). The numerical results presented in Table 3 demonstrate a rapid increase in the normalized $\chi' = \partial\chi/\partial T$ with decreasing $|J'|/J$ and show that the susceptibility at the critical point also varies in inverse proportion to $|J'|/J$.

## 6. POWDER MAGNETIC SUSCEPTIBILITY OF $CoCl_2 \cdot 2NC_5H_5$ AND $FeCl_2 \cdot 2NC_5H_5$

The spin systems of crystals of the pyridine complexes of cobalt and iron(II) chlorides are 3D Ising lattices with quasi-1D coupling [1, 2]. In these compounds, $Co^{2+}$ or $Fe^{2+}$ ions are linked by next-nearest-neighbor superexchange coupling through chlorine into linear chains separated by pyridine rings, which are responsible for interchain coupling. At temperatures on the order of the critical temperature, the only significantly populated level in the energy level system of metal ions modified by the single-ion anisotropy field is the ground-state Kramers doublet, which is separated from higher levels by a large energy gap. Therefore, the coupling between these ions can be modeled by an effective spin-1/2 Hamiltonian.

**Table 3.** Normalized coordinates of the inflection point of longitudinal susceptibility below its maximum and the values of susceptibility and its temperature derivative at the inflection point as functions of $J'/J$ for $4 \times 4 \times \infty$ lattices with $J > 0$ and $J' < 0$

| $J'/J$ | $k_B T_c/J$ | $\frac{1}{3}\chi_{\parallel,c}'^{(4\times4\times\infty)}$ | $\frac{1}{3}\chi_{\parallel,c}^{(4\times4\times\infty)}$ |
|---|---|---|---|
| −1.0000 | 1.947425 | 0.06298528 | 0.01244998 |
| −0.1000 | 0.621899 | 3.41073780 | 0.13753748 |
| −0.0100 | 0.311571 | 110.197410 | 1.39414706 |
| −0.0095 | 0.307526 | 119.120733 | 1.45608732 |
| −0.0090 | 0.303649 | 127.885372 | 1.53655666 |
| −0.0085 | 0.299877 | 139.597874 | 1.63723231 |
| −0.0080 | 0.296189 | 151.066706 | 1.76240864 |
| −0.0075 | 0.291462 | 166.471661 | 1.85697964 |
| −0.0070 | 0.287312 | 182.003965 | 2.00942967 |
| −0.0065 | 0.281839 | 205.372014 | 2.11081321 |
| −0.0060 | 0.277612 | 227.136261 | 2.33597593 |
| −0.0055 | 0.272541 | 254.505992 | 2.55608624 |
| −0.0050 | 0.267087 | 290.445604 | 2.81257312 |
| −0.0045 | 0.260747 | 336.356921 | 3.06639337 |
| −0.0040 | 0.254937 | 394.237723 | 3.50292699 |
| −0.0035 | 0.248861 | 473.481579 | 4.10957821 |
| −0.0030 | 0.241041 | 591.045463 | 4.72049719 |
| −0.0025 | 0.232527 | 745.165859 | 5.60393984 |
| −0.0020 | 0.222934 | 1017.23093 | 6.95010013 |
| −0.0015 | 0.211715 | 1510.16621 | 9.23674430 |
| −0.0010 | 0.197356 | 2567.00852 | 13.6273345 |

The data points in Fig. 8 represent the zero-field magnetic susceptibilities of polycrystalline $CoCl_2 \cdot 2NC_5H_5$ and $FeCl_2 \cdot 2NC_5H_5$ powders measured in [11] and [17], respectively.

For $CoCl_2 \cdot 2NC_5H_5$ crystals, the Curie constant is $C = 2.82(5)$ cm$^3$/mol, the Curie–Weiss temperature is $\Theta = 4.95(5)$ K, and the low-temperature limit value of powder susceptibility is $\chi_p(0) = 0.14(1)$ cm$^3$/mol [11]. These numerical values are consistent with upper bound (23). Unfortunately, inequality (22) cannot be used to obtain any useful quantitative information because of a large experimental error. However, large values of maximum susceptibility (see Fig. 8) suggest that the interchain coupling is weak. The corresponding contribution to longitudinal susceptibility can be ignored, since it is much smaller than the $\chi_p(0)$ measurement error. If the contribution of $J'$ is also ignored, then the decoupled system (19)–(21) yields $J/k_B = 11.4$ K, $g_\parallel = 6.26$, and $g_\perp = 5.05$, and only the value of $J'/J$ is required to calculate the susceptibility curve.

The critical temperature and maximum susceptibility calculated for the 3D Ising lattice as functions of $|J'|/J$ are qualitatively similar to those shown in Fig. 5 for the 2D model. Numerical values of $k_B T_c/J$ in the 3D Ising model can be found in a recent paper ([34], Table 3, the values in the $T_c$ column divided by 2). Using these values and taking $T_c = 3.17(2)$ K and $J/k_B = 10.6(6)$ K for $CoCl_2 \cdot 2NC_5H_5$ from [11], we obtain $|J'|/J = 0.0069^{+0.0022}_{-0.0016}$ with an error of 30%. Analogously, taking $T_c = 6.6(3)$ K and $J/k_B = 25(2)$ K for $FeCl_2 \cdot 2NC_5H_5$ from [17], we obtain $|J'|/J = 0.0038^{+0.0033}_{-0.0015}$; i.e., the corresponding error is even larger. Therefore, $|J'|/J$ should again be determined from the maximum susceptibility.

Figure 7 demonstrates that the transverse susceptibility in the neighborhood of the longitudinal-susceptibility maximum point is almost constant, and its value predicted by (15) combined with (A.12) is $[\chi_\perp^{(1D)}(0) + \chi_{\perp,\max}^{(1D)}]/2 \approx 0.27496$ (measured in units of $J/N_A g_\perp^2 \mu_B^2$. Accordingly, the maximum powder susceptibility of quasi-1D superantiferromagnets with $10^{-3} \leq |J'|/J \leq 10^{-2}$ can be calculated as

$$\frac{J\chi_p^{(\max)}}{N_A g_\parallel^2 \mu_B^2} \approx \frac{1}{3}\frac{J\chi_{\parallel,\max}^{(4\times4\times\infty)}}{N_A g_\parallel^2 \mu_B^2} + 0.1833\left(\frac{g_\perp}{g_\parallel}\right)^2, \quad (34)$$

with $J\chi_{\parallel,\max}^{(4\times4\times\infty)}/3N_A g_\parallel^2 \mu_B^2$ taken from Table 2.

Using the experimental value $\chi_p^{(\max)} = 3.9$ cm$^3$/mol for $CoCl_2 \cdot 2NC_5H_5$ determined from Fig. 4 in [11] and applying the $4 \times 4 \times \infty$ cluster model, we find that $|J'|/J = 6.53 \times 10^{-3}$.

Curve *1* in Fig. 8 is the powder susceptibility

$$\chi_p(T) \approx \frac{1}{3}[\chi_\parallel^{(4\times4\times\infty)}(T) + 2\chi_\perp^{(1D)}(T)] \quad (35)$$

calculated by using the parameters obtained for $CoCl_2 \cdot 2NC_5H_5$. Here, the maximum is located at 3.48 K, which agrees with $T_{\max} = 3.51(1)$ K measured in [11]. The inflection point below the maximum of the theoretical curve is located at 3.2 K. This result is also consistent with $T_c = 3.17(2)$ K determined by measuring specific heat in [11].

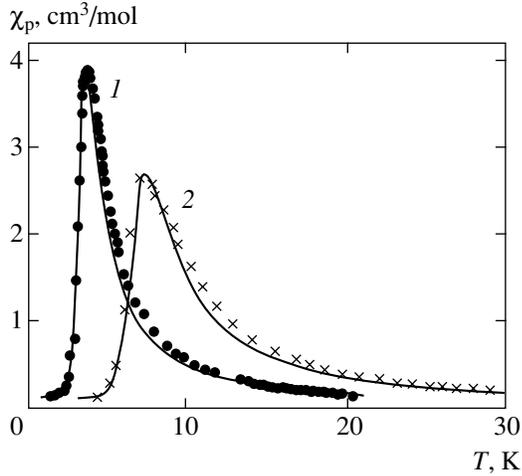

**Fig. 8.** Temperature dependence of $CoCl_2 \cdot 2NC_5H_5$ (*1*) and $FeCl_2 \cdot 2NC_5H_5$ (*2*) powder susceptibility: measurement data from [11] (circles) and [17] (crosses) and theoretical predictions based on (35) with $J/k_B = 11.4$ K, $J'/J = -6.53 \times 10^{-3}$, $g_\parallel = 6.26$, and $g_\perp = 5.05$ for $CoCl_2 \cdot 2NC_5H_5$ and $J/k_B = 24.5$ K, $J'/J = -5.65 \times 10^{-3}$, $g_\parallel = 7.07$, and $g_\perp = 6.85$ for $FeCl_2 \cdot 2NC_5H_5$ (curves).

The experimental data for $FeCl_2 \cdot 2NC_5H_5$ presented in [17] were used to obtain $C = 4.5$ cm³/mol, $\Theta = 8.5$ K, $\chi_p(0) = 0.12$ cm³/mol, and $\chi_p^{(max)} = 2.7$ cm³/mol. Again, the model parameters were adjusted to find $J/k_B = 24.5$ K, $|J'|/J = 5.65 \times 10^{-3}$, $g_\parallel = 7.07$, and $g_\perp = 6.85$. These parameters were used in (35) to obtain curve *2* in Fig. 8, for which $T_{max} = 7.2$ K and $T_c = 6.7$ K. According to [17], $J/k_B = 25(2)$ and $T_c = 6.6(3)$; i.e., the estimates for $J/k_B$ and $T_c$ obtained in this study agree with experimental data within measurement error.

## 7. CONCLUSIONS

The temperature dependence of the longitudinal and transverse zero-field susceptibilities of 2D and 3D Ising lattices with anisotropic coupling is analyzed. The analysis is based on approximations of the original lattices with ensembles of independent chain clusters that are infinitely long in the strong-coupling direction.

A detailed treatment is presented of the ferromagnetic intrachain and antiferromagnetic interchain couplings that constitute the superantiferromagnetic coupling characteristic of the modeled systems. For this coupling configuration, longitudinal susceptibility has a maximum whose value varies in inverse proportion to interchain coupling strength. An explanation is proposed for the inverse proportionality.

It is found that the relative error in the value of the maximum calculated for clusters of the same finite size $L$ is independent of lattice anisotropy.

For highly anisotropic superantiferromagnets, it is shown that the interchain coupling strength can be determined much more accurately by using the maximum value than the critical temperature, whereas the latter method is applicable to weakly anisotropic systems.

A convergence analysis of cluster expansions performed for 2D systems shows that $L \times \infty$ strips of width $L = 10$ can be used to calculate the maximum susceptibility up to an error of 0.2%. If required, clusters of larger size can be used to improve accuracy. Currently, strips of width $L \leq 10$ can be simulated on a standard PC. Simulations for widths up to $L = 16$ can be performed on supercomputers. Strips of larger size can be simulated by using the symmetry of $L \times \infty$ cylinders and representing the transfer matrix in block diagonal form.

Three-dimensional simulation is an essentially different task. The susceptibility of $L \times L \times \infty$ Ising systems with $L \leq 4$ can be calculated on a PC, whereas the $6 \times 6 \times \infty$ problem cannot be solved on any supercomputer even after the transfer matrix is reduced to a block diagonal form.

The approximation accuracy achieved in this study is sufficient for a well-founded quantitative description of the magnetic susceptibilities measured for real 2D and 3D anisotropic Ising superantiferromagnets. The present numerical results obtained are valid in the entire experimental temperature range.

The agreement achieved between theory and experiment strongly suggests that FeTAC, $CoCl_2 \cdot 2NC_5H_5$, and $FeCl_2 \cdot 2NC_5H_5$ can be very accurately treated as Ising magnets.

The accuracy of estimation of $|J'|/J$ is improved from one or two to three digits.

## ACKNOWLEDGMENTS

This work was supported by the Russian Foundation for Basic Research, project nos. 03-02-16909 and 04-03-32528.

## *APPENDIX*

The formulas given below were used in analyzing experimental data.

### *Longitudinal Susceptibility*

The zero-field longitudinal susceptibility of a 1D Ising chain is

$$\chi_\parallel^{(1D)}(T) = \frac{N_A g_\parallel^2 \mu_B^2}{4 k_B T} e^{J/k_B T}. \quad (A.1)$$

The longitudinal susceptibility of a double-chain 1D $(2 \cdot 1D)$ Ising model is expressed as follows [46, 47]

(see also [48]):

$$\chi_\parallel^{(2\cdot 1D)}(T) = \frac{N_A g_\parallel^2 \mu_B^2}{4k_B T} \frac{e^{J/k_B T}}{A\cosh(z'J'/2k_B T)}$$

$$\times \left(A + \cosh\frac{J}{k_B T}\sinh\frac{z'J'}{2k_B T}\right) \quad (A.2)$$

$$\times \left(A + \sinh\frac{J}{k_B T}\sin\frac{z'J'}{2k_B T}\right),$$

where

$$A = \left(1 + \cosh^2\frac{J}{k_B T}\sinh^2\frac{z'J'}{2k_B T}\right)^{1/2}. \quad (A.3)$$

The expression for the longitudinal susceptibility of a four-chain (4 · 1D) Ising model (a truly 3D cluster, such as $4 \times \infty$ cylinder or $2 \times 2 \times \infty$ parallelepiped) is [30]

$$\chi_\parallel^{(4\cdot 1D)}(T) = \frac{N_A g_\parallel^2 \mu_B^2}{16 k_B T}$$

$$\times \frac{(A^2 - 4e^{2y}\sinh^4 x)F + 4AGe^y \sinh x}{(A^2 - 2Ae^y \sinh 2x \cosh y + 4e^{2y}\sinh^4 x)R_1 R_2}, \quad (A.4)$$

where

$$R_{1,2} = [1 + (\sqrt{2}\cosh x \sinh y \pm \cosh y)^2]^{1/2},$$

$$A = (\sqrt{2}\cosh x \cosh y + \sinh y + R_1)$$

$$\times (\sqrt{2}\cosh x \cosh y - \sinh y + R_2),$$

$$B = 4[1 + (\sqrt{2}\cosh x \sinh y + \cosh y + R_1) \quad (A.5)$$

$$\times (\sqrt{2}\cosh x \sinh y - \cosh y + R_2)],$$

$$F = \cosh^2 y - 2\cosh^2 x \sinh^2 y + R_1 R_2 + B - 3,$$

$$G = 2\sqrt{2}(R_1 + R_2) + \cosh x \sinh y (2B - F),$$

with $x = J/k_B T$ and $y = z'J'/2k_B T$.

According to these formulas, the longitudinal susceptibility either increases indefinitely or vanishes (exponentially) as $T \longrightarrow 0$, depending on whether the zero-temperature ordered state has nonzero or zero magnetic moment. In particular, if $J > 0$ and $J' < 0$, then the longitudinal susceptibility vanishes at $T = 0$. As $T \longrightarrow \infty$, the susceptibility vanishes according to the Curie–Weiss law (9). It should be noted here that the constant parameters in the law are the correct Curie constant and Curie–Weiss temperature given by (10) and (11), respectively.

*Transverse Susceptibility*

The zero-field transverse susceptibility of a linear Ising chain is [36, 49, 50]

$$\chi_\perp^{(1D)}(T) = \frac{N_A g_\perp^2 \mu_B^2}{4J}$$

$$\times \left[\tanh\frac{J}{2k_B T} + \frac{J/2k_B T}{\cosh^2(J/2k_B T)}\right]. \quad (A.6)$$

The zero-field transverse susceptibility of a double-chain Ising model with interchain coupling strength $z'J'$ has the form [51]

$$\chi_\perp^{(2\cdot 1D)}(T) = \frac{1}{16} N_A g_\perp^2 \mu_B^2 (A_1 - B_1 G_1 \quad (A.7)$$

$$+ 2A_2 G_2 - 2B_2 G_3 + A_3 G_4 - B_3 G_5),$$

where

$$A_1 = \frac{1}{2J + z'J'}\sinh\frac{2J + z'J'}{k_B T} + \frac{1}{2J - z'J'}$$

$$\times \sinh\frac{2J - z'J'}{k_B T} + \frac{2}{z'J'}\sinh\frac{z'J'}{k_B T}, \quad (A.8)$$

$$A_2 = \frac{1}{2J + z'J'}\sinh\frac{2J + z'J'}{k_B T} - \frac{1}{2J - z'J'}\sinh\frac{2J - z'J'}{k_B T},$$

$$A_3 = \frac{4}{2J + z'J'}\sinh\frac{2J + z'J'}{k_B T} - A_1 - 2A_2,$$

$$B_1 = \frac{1}{2J + z'J'}\left(\cosh\frac{2J + z'J'}{k_B T} - 1\right)$$

$$-\frac{1}{2J - z'J'}\left(\cosh\frac{2J - z'J'}{k_B T} - 1\right) + \frac{2}{z'J'}\left(\cosh\frac{z'J'}{k_B T} - 1\right),$$

$$B_2 = \frac{1}{2J + z'J'}\left(\cosh\frac{2J + z'J'}{k_B T} - 1\right) \quad (A.9)$$

$$+ \frac{1}{2J - z'J'}\left(\cosh\frac{2J - z'J'}{k_B T} - 1\right),$$

$$B_3 = \frac{4}{2J + z'J'}\left(\cosh\frac{2J + z'J'}{k_B T} - 1\right) - B_1 - 2B_2;$$

$$G_1 = \frac{1}{R}\cosh\frac{J}{k_B T}\sinh\frac{z'J'}{2k_B T},$$

$$G_2 = \frac{1}{R}\sinh\frac{J}{k_B T}\sinh\frac{z'J'}{2k_B T},$$

$$G_3 = \frac{1}{R}\sinh\frac{J}{k_B T}\left(\cosh\frac{z'J'}{2k_B T} - \frac{1}{S}\cosh\frac{J}{k_B T}\right),$$

$$G_4 = \frac{1}{S}\sinh^2\frac{J}{k_B T}\left(\frac{2}{R}\sinh^2\frac{z'J'}{2k_B T} + \frac{1}{S}\right),$$

$$G_5 = \frac{1}{RS}\sinh^2\frac{J}{k_B T}\sinh\frac{z'J'}{2k_B T}$$
$$\times\left(2\cosh\frac{z'J'}{2k_B T} - \frac{1}{S}\cosh\frac{J}{k_B T}\right),$$

(A.10)

with

$$R = \left(1 + \cosh^2\frac{J}{k_B T}\sinh^2\frac{z'J'}{2k_B T}\right)^{1/2},$$

$$S = R + \cosh\frac{J}{k_B T}\cosh\frac{z'J'}{2k_B T}.$$

(A.11)

The transverse susceptibility is invariant under the sign changes $J \to -J$ and $J' \to -J'$. At absolute zero temperature, susceptibility (A.7) reduces to (15), which is the correct value for a system of any dimensionality. Figure 2 demonstrates that the transverse susceptibility is nearly constant at low temperatures, reaches a maximum at a higher temperature, and follows the high-temperature Curie law (13), (14).

The coordinates of the maximum of transverse susceptibility (A.6) are

$$\frac{k_B T_{max}}{|J|} = 0.418778\ldots, \quad \frac{|J|\chi_{\perp, max}^{(1D)}}{N_A g_\perp^2 \mu_B^2} = 0.299919\ldots.$$

(A.12)

For the double-chain Ising model with $z'|J'|/J| = 0.1$ (curve 2 in Fig. 2), the transverse-susceptibility maximum has the coordinates (0.480876, 0.288263).